\documentstyle[11pt,aaspp]{article}
\input{psfig}
\eqsecnum
\begin{document}
\title{Cosmological Evolution of Quasars}

\author{Insu Yi}
\affil{Institute for Advanced Study, Olden Lane, Princeton, NJ 08540;
yi@sns.ias.edu}

\begin{abstract}

We present a model for the cosmological evolution of quasars (QSOs) 
under the assumption that they are powered by massive accreting black 
holes. Accretion flows around massive black holes make a transition from high 
radiative efficiency ($\sim 10\%$) to low efficiency, advection-dominated 
flows when ${\dot M}/{\dot M}_{Edd}$ falls below the critical rate 
$\sim 0.3\alpha^2\sim 10^{-2}$ where ${\dot M}$ is the mass accretion 
rate, ${\dot M}_{Edd}\propto M$ is the usual Eddington rate with the nominal 
$10\%$ efficiency, and $\alpha (\le 1)$ is the dimensionless viscosity 
parameter. We identify this transition with the observed break at a redshift
$\sim 2$ in the QSOs' X-ray luminosity evolution. Growth of black holes 
through accretion could naturally lead to such a transition at a critical 
redshift $z_c\sim 1-3$, provided that most of high redshift QSOs appear 
with near Eddington luminosities at $z\sim 3-4$ and the accretion rates 
decline over the Hubble time in a roughly synchronous manner.
Before the transition, the QSOs' luminosities (with a high efficiency) slowly
decrease and after the transition at $z_c$, the QSO luminosities
evolve approximately as $\propto (1+z)^{K(z)}$ where $K(z)$ gradually varies 
from $z=z_c$ to $z\sim 0$ around $K\sim3$. The results depend on the details 
of the QSO X-ray emission mechanism. We discuss some further implications.

\end{abstract}


\section{Introduction}

The cosmological origin and evolution of quasars (QSOs) has remained an 
outstanding issue in astrophysics. QSOs first appear at a high redshift 
$z>4$ and their evolution appears very slow between $z\sim 3$ and 
the critical redshift $z_c\sim 2$. Below $z_c$, bright QSOs seem rapidly 
disappearing (for a review, see Hartwick \& Shade 1991). In the optical,
the evolution is often described as pure luminosity evolution
(Mathez 1976,1978, Weedman 1986) with the luminosity decreasing
as $(1+z)^K$ with $K\sim 3-4$ depending on the cosmological
model (Marshall 1985, Weedman 1986, Boyle et al. 1988, Hartwick
\& Shade 1991). In the X-ray, a similar evolution is seen although
the optical evolution appears somewhat steeper (Maccacaro et al. 1991, 
Della Ceca et al. 1992, Boyle et al. 1993, 1994, Page et al. 1996). 

In the conventional picture (for a review, see e.g. Rees 1990, Frank et al. 
1992), QSOs are powered by massive accreting black holes which
grow through accretion of gas. It remains largely 
unexplained why QSOs are suddenly quenched at $z<2$ (cf. Turner 1991, 
Fukugita \& Turner 1996). A tempting possibility that low $z$ Seyfert galaxies 
may be low luminosity remnants of QSOs is ruled out simply due to the fact 
that the former are more abundant than the latter at least by an order of 
magnitude (Huchra \& Burg 1992).
Why do QSOs suddenly turn off? What drives the evolution of the QSO luminosity 
function? What are the present remnants of bright QSOs? We address these
issues.

We assume that engines of QSOs are indeed massive accreting black holes. 
The accretion flows have often been modeled by a geometrically thin, 
optically thick disk (e.g. Frank et al. 1992, Pringle 1981) 
and the so-called blue bumps have been interpreted as emission from such disks
(e.g. Sun \& Malkan 1988, Wandel \& Petrosian 1988, Laor \& Netzer 1989, and
references therein). For the optical QSO evolution, 
some phenomenological prescriptions have been proposed (Heisler \& Ostriker 
1988, Koo \& Kron 1988, Boyle et al. 1988,1993, Pei 1995). 
For the X-ray evolution, it has been shown that the pure luminosity evolution
can account for the observed evolution (Boyle et al. 1993,1994, 
Page et al. 1996). Although there have been various models 
on the possible driving mechanism for the QSO cosmological evolution 
(e.g. Caditz et al. 1991, Small \& Blandford 1992, Haehnelt \& Rees 1993), 
a definite conclusion has not been reached.

For a typical QSO luminosity $L\sim 10^{46} erg/s$, a black hole of
mass $m\equiv M/10^8 M_{\sun}$ accreting at the Eddington rate with the nominal
radiative efficiency $\eta_n=0.1$,
\begin{equation}
{\dot M}_{Edd}=L_{Edd}/(\eta_n c^2)=4\pi G M /\eta_n \kappa_{es} c
=[1.4\times 10^{26} g/s] m,
\end{equation}
where $\kappa_{es}$ is the electron scattering opacity,
would exponentially grow on a time scale
\begin{equation}
t_{Edd}=\left|{\dot M}_{Edd}/M\right|^{-1}=[4.5\times 10^{7} yr] (\eta_n/0.1).
\end{equation}
This suggests that even when accretion-powered QSO activity is short-lived 
($<10^9 yr$), massive black holes residing in most luminous QSOs 
($L>10^{46} erg/s$) could grow to $M>10^{9} M_{\sun}$.
We suggest that this rapid growth itself may directly affect the QSOs'
evolution. The basic idea is that the massive black holes, accreting at a
scaled accretion rate ${\dot m}\equiv {\dot M}/{\dot M}_{Edd}$ below a 
certain critical rate ${\dot m}_c\sim 10^{-2}$ (Narayan \& Yi 1995b), 
emit radiation with an efficiency $\eta\propto {\dot m}\ll\eta_n$.
Although the idea of the low accretion efficiency (Rees et al. 1982)
has occasionally been mentioned, the lack of a self-consistent model has been
a major obstacle.
Since ${\dot M}_{Edd}\propto M$ increases due to accretion 
(even when ${\dot M}$ remains nearly constant), ${\dot m}\propto {\dot M}/M$ 
decreases and the QSOs' luminosity decrease could be a direct result of the 
rapid mass accretion. We suggest that the transition in accretion flow
from high efficiency ($\eta\sim \eta_n=0.1$) to low efficiency 
($\eta\ll \eta_n$) flows could naturally explain the major 
properties of the X-ray luminosity evolution of QSO population (Yi 1996).

\section{Cosmological Evolution of QSO Luminosity: Transition in Accretion Flows}

The idea of transition in accretion flow 
hinges on the recently proven stability of the advection-dominated flows
(Narayan \& Yi 1994, 1995ab, Abramowitz et al. 1995, Chen et al. 1995). 
As the accretion rate falls, the density of the accreting plasma drops to
a point where the Coulomb ion-electron energy transfer
becomes inefficient compared with the rate of viscous heating on ions 
(Rees et al. 1982). In the absence of other couplings, 
the accretion flow becomes
a two temperature plasma with a very low radiative cooling efficiency
(Narayan \& Yi 1995b, Rees 1982) and
most of accretion energy is lost into black holes through advection.
Optically thin (magnetized) advection-dominated flows emit synchrotron 
(in radio) and Comptonized synchrotron and 
bremsstrahlung emission (in X-ray) resulting in characteristically flat
spectra. Detailed spectral calculations and results have been successfully
applied to low luminosity systems such as Galactic black hole systems 
(Narayan et al. 1996), the Galactic center (Narayan et al. 1995), and the
extragalactic system NGC 4258 (Lasota et al. 1996).
In the case of the Galactic center, the estimated accretion efficiency
$\eta\sim 10^{-5}\alpha^{-1}$ and in NGC 4258
$\eta\sim 10^{-4}\alpha^{-1}$ where $\alpha\sim 10^{-2}-1$ is the
dimensionless viscosity parameter (e.g. Pringle 1981, Frank et al. 1992). 
Fabian and Rees (1995) have also suggested that dim galactic nuclei 
in large elliptical galaxies (due to low efficiency advection-dominated flows) 
contain remnants of luminous QSOs.

An accretion flow enters the low efficiency regime at a critical accretion rate, 
${\dot m}_{c}\equiv {\dot M}_{c}/{\dot M}_{Edd}$, 
which is primarily determined by $\alpha (\le 1)$ (Narayan \& Yi 1995b),
\begin{equation}
{\dot m}_{c}\approx {\dot M}_{c}/{\dot M}_{Edd}=0.3\alpha^2
\end{equation}
for $R<10^3R_s$ where $R$ is the radial distance and $R_s=2GM/c^2$. 
For $R>10^3 R_s$,
\begin{equation}
{\dot m}_{c}\approx 0.3\alpha^2(R/10^3R_s)^{-1/2}.
\end{equation}
Since most of the accretion energy is dissipated at $R<10^3 R_s$, the
critical rate given in eq. (2-1) is relevant.
When ${\dot m}>{\dot m}_{c}$, the accretion flow
takes the form of high efficiency (i.e. $\eta\sim\eta_n=0.1$) or
\begin{equation}
L=\eta_n {\dot M} c^2.
\end{equation}
For $\dot m<\dot m_{c}$,
the low efficiency is approximately described by the luminosity 
\begin{equation}
L\approx 30{\dot m}^xL_{Edd}
\end{equation}
with $x\approx 2-2.2$. We take $\alpha=0.3$ or ${\dot m}_{c}\sim 0.03$
and $x=2.2$ for our discussions (Narayan \& Yi 1995b). 
Unless the accretion rate rapidly increases toward the present epoch, 
the growth of black holes always tend to decrease
${\dot m}={\dot M}/{\dot M}_{Edd}\propto {\dot M}/M$, 
which naturally drives the transition. The overall evolution takes the
form of the the pure luminosity evolution.

The direct comparison with the observed X-ray evolution is complicated by
the uncertainties in X-ray emission from luminous QSOs. The conventional
thin disks (with high efficiency) have the effective disk temperatures 
$\le 10^6$ K and hence cannot emit X-rays efficiently (e.g. Frank et al. 1992).
We assume that a significant fraction of the bolometric
luminosity is emitted as X-rays (presumably from accretion disk coronae) and
that $\eta_n=0.1$ represents this uncertainty.
The advection-dominated hot flows differ from the thin disks as the former
emit hard X-rays up to several $100$ keV. 
As long as most of the accretion flow is advection-dominated, 
the emission spectra are characteristically flat, i.e. luminosity
per decade of frequency is roughly constant (e.g. Narayan et al. 1995,
Lasota et al. 1996). The X-ray luminosity can be approximated as a fixed
fraction of the total luminosity. In our discussions, we simply assume that the 
observed X-ray luminosity is proportional to the bolometric luminosity.
We caution that our assumption could introduce serious errors if the
X-ray to bolometric luminosity ratio is a complex function of some yet
unknown parameters such as $L$, ${\dot M}$, or ${\dot m}$.
For instance, if the accretion flow contains a thin disk
in the outer region (eq. (2-2)), the resulting 
emission spectra could be significantly affected. In this sense, it is
uncertain how our suggestion on the X-ray luminosity evolution could
be applied to the optical evolution (see below).

We begin with a simplest scenario; 
(i) The black holes are initially accreting at their 
Eddington rates (cf. Padovani 1989). (ii) The accretion rates are 
sufficiently slowly
decreasing from the initial epoch at cosmic time $t=t_i$ (or $z=z_i$).
For a flat universe ($\Omega_0=1$) with no cosmological constant 
($\Omega_{\Lambda}=0$), the cosmic time $t$ at $z$ is given by
\begin{equation}
H_o(t-t_i)={2\over 3}\left[(1+z)^{-3/2}-(1+z_i)^{-3/2}\right]
\end{equation}
where $t_i=t(z=z_i)$. When ${\dot M}\approx constant$, 
or
\begin{equation}
M\approx M_i+{\dot M}(t-t_i)
\end{equation}
where $M_i=M(z=z_i)$,
\begin{equation}
{\dot m}\propto M^{-1}\propto \left[1+{2{\dot M}\over 3M_i H_o}\left((1+z)^{-3/2}
-(1+z_i)^{-3/2}\right)\right]^{-1}.
\end{equation}
where $H_o=50km/s/Mpc$ is the assumed Hubble constant.
While ${\dot m}>{\dot m}_c$, the observed slow luminosity evolution in X-rays 
and optical could be purely due to the gradual decrease of ${\dot M}$ 
(eq. (2-3)). When ${\dot m}<{\dot m}_c$, the luminosity evolves as (eq. (2-4))
\begin{equation}
L\propto \left[1+{2{\dot M}\over 3M_i H_o}\left((1+z)^{-3/2}
-(1+z_i)^{-3/2}\right)\right]^{-x+1}.
\end{equation}
In the limit that the accreted mass over Hubble time is larger than the
initial mass, i.e. ${\dot M}/H_o\gg M_i$,
\begin{equation}
L\propto (1+z)^{3(x-1)/2}\left[(1+z_i)^{3/2}-(1+z)^{3/2}\right]^{-x+1}
\end{equation}
Similarly, for an empty universe ($\Omega_0=\Omega_{\Lambda}=0$),
\begin{equation}
H_o(t-t_i)=\left[(1+z)^{-1/2}-(1+z_i)^{-1/2}\right]
\end{equation}
and
\begin{equation}
L\propto \left[1+{{\dot M}\over M_i H_o}\left((1+z)^{-1/2}
-(1+z_i)^{-1/2}\right)\right]^{-x+1}.
\end{equation}
or in the limit ${\dot M}/H_o\gg M_i$
\begin{equation}
L\propto (1+z)^{(x-1)/2}\left[(1+z_i)^{1/2}-(1+z)^{1/2}\right]^{-x+1}
\end{equation}
The simple behavior of luminosity as a function of redshift is often
approximated as
\begin{equation}
L(z)\propto (1+z)^K,
\end{equation}
with a constant $K$. In our scenario, $K(z)$ is a function of $z$ \& $z_i$ and 
depends on the cosmological model. We get for the flat universe,
\begin{equation}
K(z)={d\ln L(z)\over d\ln (1+z)}={3(x-1)\over 2}\left[1+{1\over 
((1+z_i)/(1+z))^{3/2}-1}\right]
\end{equation}
and for the empty universe,
\begin{equation}
K(z)={x-1\over 2}\left[1+{1\over ((1+z_i)/(1+z))^{1/2}-1}\right]
\end{equation}
where eqs. (2-9),(2-12) have been used. The redshift dependence of $K$ is
shown in Fig. 1.
In the two cosmological models considered, which cover various other possible
models, the evolution at $z<z_i$ is shown to give a wide range
of $K$ which includes $K\sim 2-3$. $K$ varies from $K>3$ to $K<3$ as $z$ 
decreases away from $z_i$. The derived expressions for $K(z)$ become invalid
for $z\rightarrow z_i$ as the accretion flows become those with a high 
efficiency in our scenario. The derived values of $K$ are interestingly
close to those obtained by Boyle et al. (1994) and Page et al. (1996)
using the recent X-ray observations (see below).

Another important aspect of the QSO evolution is the sudden cutoff in the
luminosity evolution near $z=z_c\sim 2$ which we identify as the critical
redshift at which $\dot m=\dot m_c$ first occurs. 
For the flat universe with no cosmological constant, we get
\begin{equation}
{\dot m}_c=0.3\alpha^2={\eta_{n} \kappa_{es} c\over 4\pi G}{\dot M\over M_i}
\left[1+{2{\dot M}\over 3H_o M_i}\left[(1+z_c)^{-3/2}-(1+z_i)^{-3/2}\right]
\right]^{-1}.
\end{equation}
If the initial accretion rate is a fraction $\delta(\le 1)$ of the initial 
Eddington rate or
\begin{equation}
{\dot M}/M_i=\delta t_{Edd}^{-1}= (4.5\times 10^7 yr)^{-1} (\eta_{n}/0.1)^{-1}
\delta,
\end{equation}
\begin{equation}
1+z_c=\left[\left(t_{Edd}\over t_{age}\right)\left({1\over {\dot m}_c}
-{1\over\delta}\right)+(1+z_i)^{-3/2}\right]^{-2/3}
\end{equation}
where $t_{age}=2t_H/3=2/3H_o$ is the age of the Universe. 
Similarly, for the empty universe ($\Omega_0=\Omega_{\Lambda}=0$)
\begin{equation}
1+z_c=\left[\left(t_{Edd}\over t_{age}\right)\left({1\over 
{\dot m}_c}-{1\over\delta} \right)+(1+z_i)^{-1/2}\right]^{-2}
\end{equation}
where $t_{age}=t_H=1/H_o$. 
Unless $\delta\le {\dot m}_c\sim 0.03$ (which is unlikely
given the high luminosities of QSOs at high redshift $z=z_i$ and a relatively
short time scale allowed for growth of black holes from $z>z_i$),
$z_c$ only weakly depends on $\delta\sim 1$. Assuming $\alpha=0.3$,
$\delta=1$, and $H_o=50km/s/Mpc$, we get $z_c\approx 1.6 (1.8)$ for
$z_i=3 (4)$ in the flat universe and $z_c\approx 1.9 (2.6)$ for $z_i=3 (4)$ 
in the empty universe.
Increasing $H_o$ to $H_o=65km/s/Mpc$ would change $z_c$ roughly by 
$\sim 10-20$\%. For ${\dot m}_c\ll \sim 3\times 10^{-2}$ or $\alpha<0.3$, 
however, $z_c$ could be much lower than the observed $z_c\sim 2$
unless the initial accretion rate is already significantly sub-Eddington.
Our proposed evolutionary model suggests a possible explanation for the 
cutoff at $z_c\sim 2$. A definite final answer requires fine details of the 
cosmological model and the initial conditions. Nevertheless, it remains
valid that the transition driven by mass growth could always result in
rapid evolution of QSOs regardless of any specific cosmological scenario
such as the present one.

Now we consider luminosity evolution in general cosmological models relaxing
${\dot M}=constant$. We adopt the mass accretion rate of the form
\begin{equation}
{\dot M}={\dot M}_i\exp\left(-(t-t_i)/t_{evol}\right)
\end{equation}
with the evolutionary time scale as $t_{evol}=\beta t_{age}$
($\beta\le 1$). In order to see the effects of the cosmological model, 
we consider four different cosmological models (e.g. Kolb \& Turner 1990);
(i) an open universe $\Omega_0=0.4$, $\Omega_{\Lambda}=0$ for which
$t_{age}=0.779(1/H_o)$ and
$$
t(z)={1\over 2H_o}{\Omega_0\over (1-\Omega_0)^{3/2}}
\qquad\qquad\qquad\qquad\qquad\qquad\qquad\qquad\qquad\qquad\qquad
\qquad\qquad\qquad\qquad
$$
\begin{equation}
\times\left[{2(1-\Omega_0)^{1/2}(1+\Omega_0z)^{1/2}\over\Omega_0(1+z)}-\ln\left[
\Omega_0(1+z)+2(1-\Omega_0)+2(1-\Omega_0)^{1/2}(1+\Omega_0z)^{1/2}\over
\Omega_0(1+z)\right]\right],
\end{equation}
(ii) an empty universe $\Omega_0=\Omega_{\Lambda}=0$ (eq. (2-10)),
(iii) a flat universe $\Omega_0=1$, $\Omega_{\Lambda}=0$ (eq. (2-5)),
(iv) a flat universe $\Omega_0=0.4$, $\Omega_{\Lambda}=0.6$ for which
$t_{age}=0.888(1/H_o)$ and
\begin{equation}
t(z)={2\over 3H_o}{1\over (1-\Omega_0)^{1/2}}\ln\left[\left(1-\Omega_0\over
\Omega_0\right)^{1/2}\left(1\over 1+z\right)^{3/2}+\left(\left(1-\Omega_0\over
\Omega_0\right)\left(1\over 1+z\right)^3+1\right)^{1/2}\right].
\end{equation}
In each cosmological model, we need to specify initial conditions for QSO
evolution. Our simple initial conditions could be partly motivated by the 
optical observation that the QSO luminosity function is already established 
by $z\sim 3-4$ and changes little from $z\sim 3$ to $z\sim 2$ 
(Hartwick \& Shade 1990). The present model simply posits that the massive 
black holes arrive at the peak of their activities ($z\sim 3-4$), in a nearly 
synchronous manner, accreting from surrounding gas at a rate close to 
${\dot M}_{Edd}$.

In Fig. 2, we show evolution of QSOs in the four cosmological
models marked by numbers. We have taken $z_i=3.5$ in this example. 
In our model, the luminosity evolution is obviously scale-free for our 
chosen initial conditions.
That is, the luminosities of QSOs are simply determined by the black hole 
masses with an identical redshift dependence. The most significant uncertainty 
is the possible wide scatter in the initial conditions (i.e. $z_i$, $\delta$,
$t_{evol}$, etc.).
At $z>z_c$, the evolution is slow (eqs. (2-3),(2-20)) and the time scale of 
evolution is set by $t_{evol}$ as $\eta_n=0.1$ is maintained. 
After the transition at $z=z_c$, the difference in the power-law among the 
shown models is small. 
It is encouraging that the observed constant $K\sim 3$ between
$z=0$ and $z=z_c\sim 2$ could be accounted for with small effects from the 
cosmological model. For the same set of parameters and the initial conditions, 
$z_c$ significantly increases from the flat universe to the empty universe. 
Such an effect is, however, indistinguishable from that due to uncertainties
in ${\dot m}_c$ or $\alpha$. $t_{evol}\ll t_{age}$ or $\beta\ll
0.5$ could result in much faster evolution than the observed.
A faster decrease of ${\dot M}$ near $z=0$ would inevitably lead to the 
further acceleration of evolution which might be directly observable as 
deviation from the observed power-law evolution with a constant $K$.

X-ray observations have established a well defined evolutionary 
trend although details differ in several analyses. Della Ceca 
et al. (1992) and Maccacaro et al. (1991) obtained $K=2.56$ and Boyle
et al. (1993) derived $K=2.75-2.80$ for the empty universe and
$K=2.76-2.73$ with $z_c\sim 2$ for the flat universe with no cosmological
constant. More recently, Boyle et al. (1994) analyzed a larger sample
and re-derived $K=3.34$ ($z_c=1.79$) for the empty universe and
$K=3.25$ ($z_c=1.60$) or $K=3.17$ ($z_c=1.48$) for the flat universe.
Page et al. (1996) note that no evolution for $z>1.8$ is entirely compatible
with the observed evolution with $K=2.35-2.94$ \& $z_c=1.41-1.82$ 
depending on the cosmological model. Despite lacking details of evolution, 
our proposed model is largely consistent with the observations. The X-ray 
evolution is similar to the optical evolution except that the latter is 
slightly faster than (cf. Boyle et al. 1988, Boyle et al. 1994). 
In the optical, the derived $K$ depends rather sensitively on the assumed 
spectral slope which could be a major source of uncertainty (see below).

Caditz et al. (1991) has suggested that optical evolution could be
explained by the accretion disk model where the monochromatic luminosity
at a given frequency evolves rapidly as the fixed frequency crosses the Wien
cutoff frequency due to the increase of the black hole mass through
accretion. Interestingly, such an explanation also results in a luminosity
evolution of the form $L\propto {\dot m}^2$ (cf. eq. (2-4)) in the optical.
In other words, the observed $(1+z)^K$ ($K\sim 3-4$) luminosity evolution
could be the result of $L\propto {\dot m}^2$ arising from the two entirely
different origins. In our model, the long wavelngth emission is unclear due
to the uncertain nature of the outer region of the flow where both
a thin disk and an advection-dominated flow are allowed in principle. 
For instance, according to eqs. (2-1),(2-2), there could be an outer
thin disk while the inner region is fully advection-dominated when
\begin{equation}
0.3\alpha^2(R/10^3R_s)^{-1/2}<{\dot m}(R) <0.3\alpha^2
\end{equation}
at a distance $R>10^3R_s$. 
If the outer disk continues to exist at $z<z_c$, we can draw a rough estimate 
on the bolometric luminosity evolution at long wavelengths by directly 
applying eq. (2-2).
Unless the outer disk extends to a region well inside $\sim 10^3R_s$, however, 
the optical luminosity is expected to be very low and the following evolutionary
trend is not directly applicable to brightest QSOs.
As ${\dot m}\propto {\dot M}/M$ decreases, the inner region of the disk first 
becomes advection-dominated while the outer thin disk, emitting in the optical,
is maintained.
As ${\dot m}$ drops further, a larger fraction of the inner disk becomes 
advection-dominated and the thin disk is pushed outward. The
inner radius of the outer thin disk is
\begin{equation}
0.3\alpha^2\left(R_{in}\over 10^3 R_s\right)^{-1/2}={\dot m}\propto {1\over M}
\end{equation}
or
\begin{equation}
R_{in}\approx [2.7\times 10^{13} cm] m {\dot m}^{-2}
\end{equation}
for ${\dot M}\sim constant$ and $\alpha=0.3$ as before. 
The bolometric disk luminosity from the outer disk would decrease as
\begin{equation}
L_{disk}\sim {GM{\dot M}\over R_{in}}\propto {M\over R_{in}}\propto 
{M\over M^2R_s}\propto M^{-2}
\end{equation}
as $M$ increases. 

The disk emission would have the evolving peak temperature
\begin{equation}
T_{disk}\sim \left(M{\dot M}\over R_{in}^3\right)^{1/4}\propto M^{-2}\propto 
L_{disk}.
\end{equation}
or
\begin{equation}
T_{disk}\approx [2.6\times 10^3K]\left(m\over 0.1\right)^{-1/4}\left({\dot m}
\over 0.03\right)^{7/4}.
\end{equation}
The peak wavelength is
\begin{equation}
\lambda_{p}\approx [1.9\times 10^4{\AA}]\left(m\over 0.1\right)^{1/4}
\left({\dot m}\over 0.03\right)^{-7/4}\approx [2\times 10^5{\AA}]
\left(M\over 10^{8}\right)^{1/4}\left(M/M_i\over 3\times 10^2\right)^{7/4}.
\end{equation}
The emission from the outer disk is mostly in the infrared regime and the optical
emission from the disk will occur in the Wien tail of the emission at 
$\sim T_{disk}$. The monochromatic disk luminosity at a fixed optical wavelength 
$\lambda_{opt}<\lambda_p$ would rapidly decline as $\lambda_{p}\propto 
M^2$ increases and $\lambda_{opt}\ll \lambda_{p}$. In this case,
the observed optical-UV
emission is dominated by that from the inner advection-dominated flow and
the optical-UV evolution is similar to the X-ray evolution.
At a longer infrared wavelength $\lambda_{IR}>\lambda_p$, the monochromatic
luminosity 
\begin{equation}
L_{IR}\propto T_{disk}/\lambda_{IR}^2\propto M^4
\end{equation}
where we have used the Rayleigh-Jeans law. In other words, the infrared
luminosity at $\lambda_{IR}\gg \lambda_p$ would increase as $\propto M^4$
until $\lambda_{IR}\sim \lambda_p$ is reached. When $\lambda_p>\lambda_{IR}$
occurs subsequently, the infrared luminosity also rapidly declines.
One of the major uncertainties in this qualitative prediction is the effects
of dust and irradiation from the central X-ray emission region which could
seriously affect the long wavelength emission from the outer region.

\section{Implications}

If the QSO evolution is indeed due to a single population,
the entire QSO population shifts to a mass range $\ge 10^9M_{\odot}$ by $z=0$,
which suggest that massive elliptical galaxies are most likely hosts for QSO 
remnants (Fabian \& Canizares 1988, Fabian \& Rees 1995).
Such massive black holes have not been clearly identified with an exception
of M87 (Kormendy and Richstone 1996). When the radiative efficiency is
a sensitive function of the mass accretion rate, the mass estimates based 
on $\eta_n\sim 0.1$ could be seriously misleading (Narayan \& Yi 1995b, 
Fabian \& Rees 1995). In our model, the mass of a black hole at $z=0$, 
based on its luminosity, could be underestimated by more than two 
orders of magnitude. If such massive black holes reside in galactic nuclei 
at the present epoch, they may be distinguishable from lower mass black holes 
accreting at $\dot m>\dot m_{c}$ (e.g. Seyfert galaxies), primarily
due to their hard X-rays and synchrotron radio emission. 

The proposed model could be roughly compatible with the observed 
X-ray luminosity evolution if (i) the transition in accretion flow occurs at 
$L=10^{-2}L_{Edd}$ or ${\dot m}={\dot m}_c=0.3\alpha^2\sim 10^{-2}$ or
$\alpha\sim 0.3$, (ii) $\eta_n=0.1$ before transition and 
$L\propto {\dot m}^2$ after transition, and (iii) QSO remnants have 
black holes with $M>10^9M_{\odot}$. These requirements are based on our
particular set of initial conditions. For a more realistic model, some
scatter in $z_i$ and/or $\delta$ is inevitable, for which we lack
detailed information. Despite these serious uncertainties,
the simple model demonstrates a natural evolutionary trend which
essentially needs only the second point (independent of our specific proposal) 
both in optical and X-ray.

If QSOs at $z\sim 2-3$ are powered by black holes accreting at near Eddington 
rates, our explanation indicates $\dot m_{c}\sim$ a few $\times 10^{-2}$ or 
$\alpha\sim O(0.1)$ where the latter value is interestingly close to those
often quoted in Galactic accretion systems (Yi 1996, Narayan et al. 1996).
The critical accretion rate in our model is related to several theoretical
problems including the necessity for the two temperature plasma 
(Narayan \& Yi 1995a, Rees et al. 1982, Phinney 1981), which merits further 
investigations.

Alternatively, QSOs could be driven by electromagnetic power
extracted from rapidly spinning black holes (Blandford \& Znajek 1978).
Although the spin-up of massive black holes from their formation is not
clearly understood (cf. Thorne 1974), the spin-down of black holes does
provide a time scale comparable to the QSOs' cosmological evolution time
scale (Park \& Vishniac 1988) when ${\dot M}\sim {\dot M}_{Edd}$.
Such an evolutionary possibility still lacks an explanation 
as to why the accretion luminosity which is comparable to the electromagnetic 
power (Park \& Vishniac 1994) also decreases on a similar time scale.

\acknowledgments
The author thanks Masataka Fukugita, Ramesh Narayan, Martin Rees, Ed Turner, 
and Ethan Vishniac for useful discussions and comments. 
The author acknowledges support from SUAM Foundation.

\clearpage

\begin{figure}[t]
\centerline{\psfig{figure=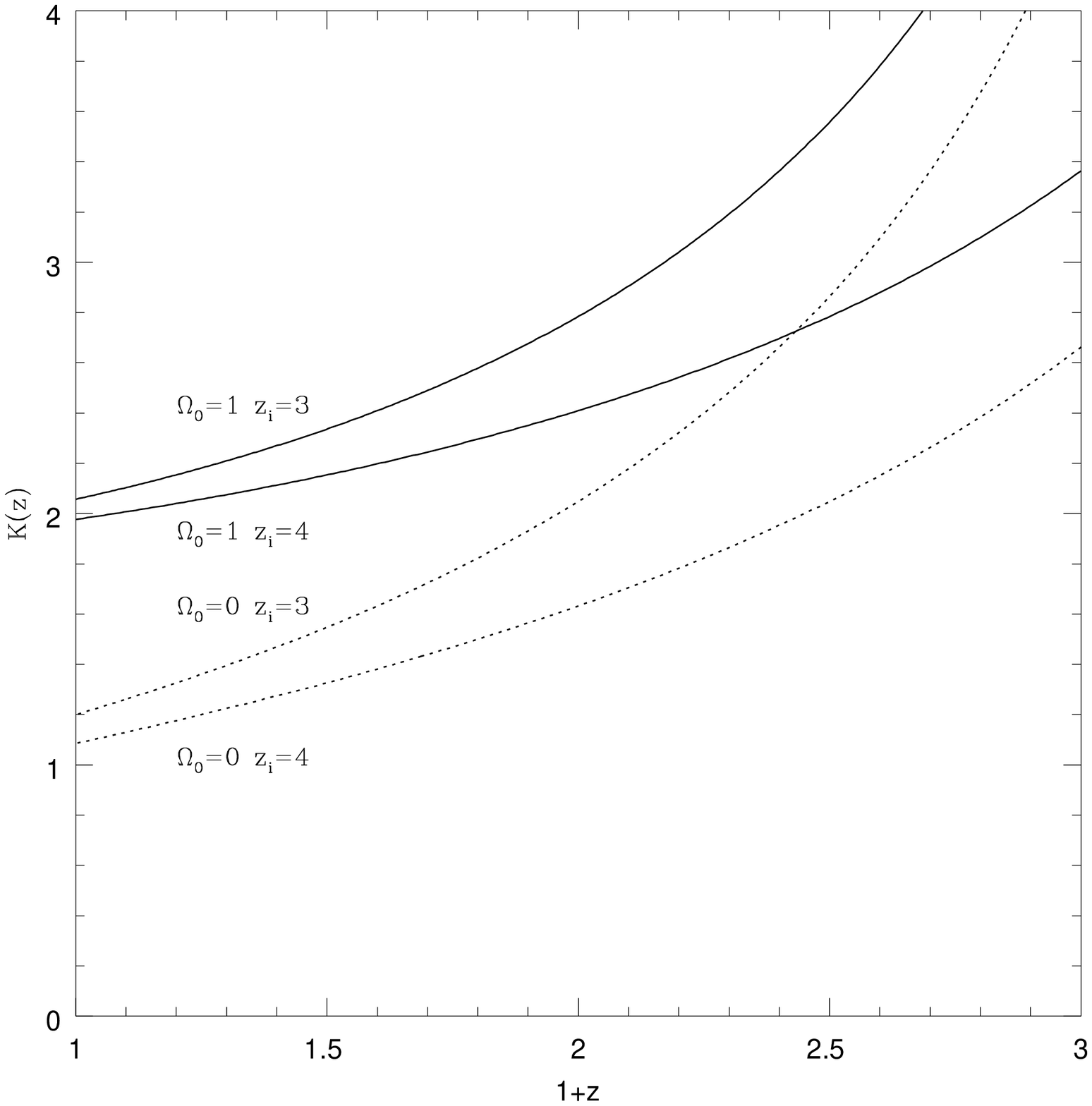,width=10.0cm,height=10.0cm}}
\caption[]
{
Luminosity evolution index K(z) in the two cosmological models 
with a constant mass accretion rate for two different 
initial epochs ($z_i=3,4$). The curves, marked by the relevant parameters,
describe the luminosity evolution for $z<z_i$ and show effects of cosmological 
model and $z_i$. Evolutions at $z<z_i$ show a wide range of $K$ including
$K\sim 2-3$ with $z$-dependence. The slope is steeper for the flat universe 
($\Omega_0=1$). $K(z)$ becomes invalid as $z\rightarrow z_i$ as the accretion 
flow is expected to enter the high efficiency regime (see text).
}
\label{fig1}
\end{figure}

\begin{figure}[t]
\centerline{\psfig{figure=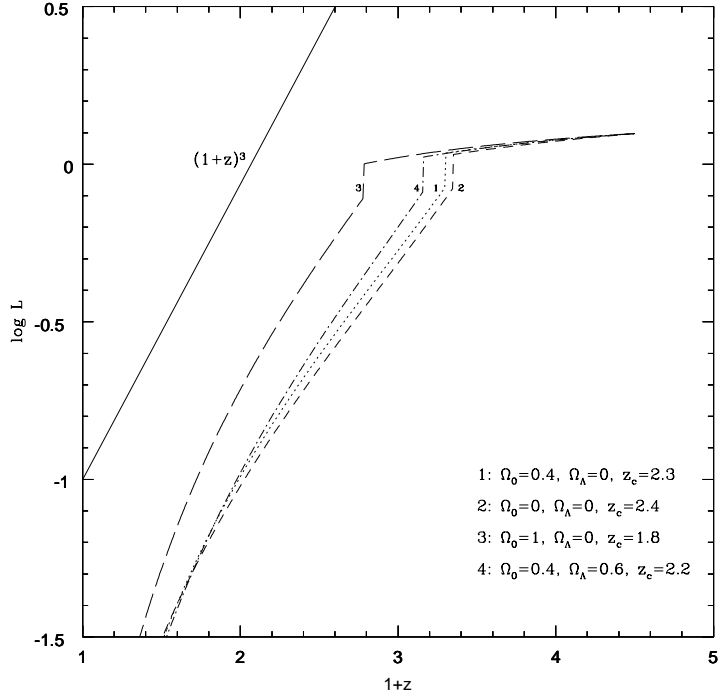,width=10.0cm,height=10.0cm}}
\caption[]
{
Scale-free luminosity evolution of a QSO in the four cosmological models 
described in the text. The cosmological model corresponding to each 
curve is marked by the number and the critical redshift $z_c$ is shown
in the figure. In this example, $\alpha=0.3$, $\beta=0.5$, and $z_i=3.5$.
For comparison, $L(z)\propto (1+z)^3$ is shown (solid line). The evolution 
becomes steeper than $(1+z)^3$ at $z<0.5$ as the mass accretion rate starts
to reflect the exponential factor on a time scale $t_{evol}=\beta
t_{age}$ with $\beta=0.5$.
}
\label{fig2}
\end{figure}

\end{document}